\documentclass[prl,amsmath,aps,10pt,letterpaper,tightenlines,twocolumn]{revtex4}
\usepackage{bm}
\usepackage{epstopdf}
\usepackage{stmaryrd}
\usepackage{amsmath}
\usepackage{calligra}
\usepackage{mathtools}
\usepackage{amsfonts}
\usepackage{amssymb}
\usepackage{textcomp}
\usepackage{graphicx}
\usepackage{caption}
\usepackage[margin=15mm]{geometry}
\usepackage{yfonts}
\usepackage[breaklinks=true,colorlinks=true,linkcolor=blue,urlcolor=blue,citecolor=blue]{hyperref}
\usepackage[lofdepth,lotdepth]{subfig}
\usepackage{color}

\newcommand{\p}{{\partial}}
\newcommand{\pt}{{\partial_t}}

\newcommand{\curl}{{\nabla\times}}
\newcommand{\haf}{{\frac{1}{2}}}
\newcommand{\intf}{{\int_0^{\infty}\,}}
\newcommand{\la}{{\langle}}
\newcommand{\ra}{{\rangle}}
\begin{document}
\title{The radiative heat transfer between a rotating nanoparticle and a plane surface}
\author{Vahid Ameri$^{1,2}$}
\email{vahameri@gmail.com}
\author{Mehdi Shafiei Aporvari$^{1}$}
\author{Fardin Kheirandish$^{1}$}
\email{fkheirandish@yahoo.com}

\affiliation{$^1$Department of Physics, Faculty of Science, University of Isfahan, Isfahan, Iran\\
$^2$ Department of Physics, Faculty of Science, University of Hormozgan, Bandar-Abbas, Iran }

\begin{abstract}
Based on a microscopic approach, we propose a Lagrangian for the combined system of a rotating dielectric nanoparticle above a plane surface in the presence of electromagnetic vacuum fluctuations. In the framework of canonical quantization, the electromagnetic vacuum field is quantized in the presence of dielectric fields describing the nanoparticle and a semi-infinite dielectric with planar interface. The radiative heat power absorbed by the rotating nanoparticle is obtained and the result is in agreement with previous results when the the rotational frequency $\omega_0$ of the nanoparticle is zero or much smaller than the relaxation frequency of the dielectrics. The well known near field effect is reexamined and discussed in terms of the rotational frequency. The radiative heat power absorbed by the nanoparticle for well-known peak frequencies, is plotted in terms of the rotational frequency $\omega_0$ showing an interesting effect resembling a phase transition around a critical frequency $\Gamma$, determined by the relaxation frequency of the dielectrics.
\end{abstract}
\maketitle
Recent developments of nanotechnology and nanoscale devices like scanning thermal microscopes \cite{abramson1999recent,kittel2005near,taubner2006near} have raised a lot of questions in the well-known category of near field effects. Radiative Heat Transfer (RHT) between two objects, like a tip of microscope near to a substrate, \cite{esteban2007tip,hartschuh2008tip} is one of these challenging problems. This problem has been studied in some special cases like two semi-infinite bodies \cite{PhysRevB.77.035431,PhysRevLett.107.014301} and a point like particle close to a plane interface \cite{mulet2001nanoscale,PhysRevB.78.115303}. Since the particle radius, in this case, is much smaller than the Wien wavelength of thermal radiation, the spectra of radiation differs significantly from the macroscopic body radiation like black body \cite{dedkov2014radiation}.
\par While investigating these static nanoscale effects have attracted a lot of interests, adding motion and studying them dynamically, where it could raise some interesting quantum effects, like quantum friction \cite{weiss1999quantum,pendry2010quantum} and dynamical Casimir effect \cite{milton2001casimir}, is more challenging.
\par The aim of the present work is to drive the radiative heat power absorbed by a dielectric nanoparticle rotating along it's axis of symmetry above a plane interface (FIG. 1), in the framework of the canonical field quantization approach. The approach is a generalization of the scheme introduced in \cite{kheirandish2009canonical}. For this purpose, we first find the explicit form of the quantized electromagnetic and also dielectric fields using Heisenberg equations of motion in non-relativistic regime (velocity of a point on the surface of the nanoparticle is much less than the speed of light), then the derivation of the radiation heat power absorbed by the nanoparticle is a straightforward problem in this scheme. A general formula for the radiation heat absorbed by a rotating nanoparticle above a plane surface is obtained and the effect of rotation on the absorbed radiation heat is studied. A feature guaranteeing the consistency of the approach, is the coincidence of the results obtained here with the previous results \cite{mulet2001nanoscale, shchegrov2000near} for the case $\omega_0=0$ or when $\omega_0 \ll \Gamma$, where $\omega_0$ and $\Gamma (1/\tau)$ are the angular velocity of the rotation and relaxation frequency(time) respectively. Actually, in this case the rotating and non-rotating particle have the same spectrum.
\par The Lagrangian describing the whole system is the Lagrangian of the electromagnetic vacuum field plus terms modeling the dielectrics and their interaction with the electromagnetic vacuum field by continuum of harmonic oscillators \cite{huttner1992quantization,kheirandish2014electromagnetic}. The rotating nanoparticle and the semi-infinite bulk interact with the electromagnetic vacuum field and the radiation heat transfer happens during this indirect interaction. The radiative heat transfer for a static point like particle has been studied widely \cite{mulet2001nanoscale,PhysRevB.78.115303,dedkov2008vacuum}. Following the method introduced in \cite{kheirandish2014electromagnetic}, we study the heat transferred to the rotating nanoparticle and its physical consequences.
\par Let us consider the following Lagrangian for a spherical nanoparticle rotating along its symmetric axis (z-axis) with angular velocity $\omega_0$, in the vicinity of a semi-infinite dielectric space
\begin{eqnarray}\label{L}
\mathcal{L} &=& \haf\epsilon_0\,(\pt \mathbf{A})^2-\frac{1}{2\mu_0}(\curl\mathbf{A})^2+\haf\intf d\nu \,[(\pt \mathbf{X}\\ \nonumber
&+&\omega_0\p_{\varphi}\mathbf{X})^2-\nu^2\mathbf{X}^2]
- \epsilon_0\intf d\nu\,f_{ij}(\nu,t)X^j\pt A_i \\ \nonumber
&+&\epsilon_0\intf d\nu\,f_{ij}(\nu,t)X^j (\mathbf{v}\times\curl\mathbf{A})_i
+\haf\intf d\nu \,[(\pt \mathbf{Y})^2\\ \nonumber
&-&\nu^2\mathbf{Y}^2]
- \epsilon_0\intf d\nu\,g_{ij}(\nu)Y^j\pt A_i,
\end{eqnarray}
where $X_i $ and $ Y_i$ are the dielectric fields describing the nanoparticle and the bulk dielectric, respectively. The bulk considered to be in local thermodynamical equilibrium at temperature $T$. To drive this Lagrangian, the coordinate-derivative and field transformations between rotating and fixed or laboratory frames are used. Note that in the laboratory or fixed frame, the electromagnetic fields are non-rotating fields, therefore, there is no need to modify them in Eq.(\ref{L}). The nanoparticle is moving so the coupling tensor $f_{ij}$ between the vacuum field and the field $X_i$ should be time dependent while the coupling tensor $g_{ij}$ between the vacuum field and the field $Y_i$, describing the non-moving bulk dielectric, is time independent.
\begin{eqnarray}\label{C}
&&f_{ij} (\nu,t)=\left(
  \begin{array}{ccc}
    f_{xx} (\nu) \cos(\omega_0 t) & f_{xx} (\nu) \sin(\omega_0 t) & 0 \\
    -f_{yy} (\nu) \sin(\omega_0 t) & f_{yy} (\nu) \cos(\omega_0 t) & 0 \\
    0 & 0 & f_{zz}(\nu)\\
  \end{array}
\right),  \nonumber \\
&&g_{ij} (\nu)=\left(
  \begin{array}{ccc}
    g_{xx} (\nu) & 0 & 0 \\
    0 & g_{yy} (\nu) & 0 \\
    0 & 0 & g_{zz}(\nu)\\
  \end{array}
\right).
\end{eqnarray}
Here we are considering homogeneous dielectrics where the coupling tensors inside them are position independent and outside the dielectrics are identically zero. We also assuming that the local velocity $\mathbf{v}=\mathbf{\omega_0}\times\mathbf{r}$ (the velocity of a point with position $\mathbf{r}$ in the particle frame) is non-relativistic, i.e., $\vert \mathbf{v} \vert \ll c$, so we can ignore from the magnetic term in Eq.(\ref{L}).
To canonically quantize the system, we need the conjugate momenta corresponding to the fields. From Eq.(\ref{L}) we have
\begin{eqnarray}\label{Q}
\Pi_i (\mathbf{r},t)&=&\frac{\p \mathcal{L}}{\p (\pt A_i)}=-\epsilon_0 E_i-P_{P,i} -P_{B,i}, \nonumber\\ 
Q^P_i&=&\frac{\p \mathcal{L}}{\p (\pt X_i)}=\pt X_i+\omega_0\,\p_{\varphi}X_i,\nonumber\\
Q^B_i&=&\frac{\p \mathcal{L}}{\p (\pt Y_i)}=\pt Y_i,
\end{eqnarray}
where we have defined $P_{P,i} (\mathbf{r},t)=\epsilon_0\intf d\nu f_{ij} (\nu,t)\,X^j (\mathbf{r},t,\nu)$ and $P_{B,i} (\mathbf{r},t)=\epsilon_0\intf d\nu g_{ij} (\nu)\,Y^j (\mathbf{r},t,\nu)$ as the electric polarization components of the nanoparticle and the bulk dielectric respectively. The system is quantized by imposing the equal-time commutation relations
\begin{eqnarray}\label{Q1}
[A_i (\mathbf{r},t),\Pi_j (\mathbf{r}',t)]&=&i\hbar\,\delta_{ij}\,\delta (\mathbf{r}-\mathbf{r}'),\\ \nonumber
[X_{i} (\mathbf{r},t,\nu),Q^P_{j} (\mathbf{r}',t,\nu')]&=&i\hbar\,\delta_{ij}\,\delta (\mathbf{r}-\mathbf{r}')
\delta (\nu-\nu'), \\ \nonumber
[Y_{i} (\mathbf{r},t,\nu),Q^B_{j} (\mathbf{r}',t,\nu')]&=&i\hbar\,\delta_{ij}\,\delta (\mathbf{r}-\mathbf{r}')
\delta (\nu-\nu').
\end{eqnarray}
The equations of motion for the electromagnetic and matter fields can be obtained from Euler-Lagrange equations or equivalently from Hamiltonian, using Heisenberg equations. By making use of the azimuthal symmetry and non-relativistic approximation, we find \cite{kheirandish2014electromagnetic}
\begin{eqnarray}\label{FE}
\mathbf{P}_P (\mathbf{r},\omega)&=&\mathbf{P}_P^{N} (\mathbf{r},\omega)+\epsilon_0 \boldsymbol{\chi}^{P} (\omega,-i\p_{\varphi})\mathbf{E}, \nonumber \\
\Bigl\{\curl\curl &-&\frac{\omega^2}{c^2}\mathbb{I}-\frac{\omega^2}{c^2}\boldsymbol{\chi}^{P}(\omega,-i \p_\varphi) -\frac{\omega^2}{c^2}\boldsymbol{\chi}^{B}(\omega) \Bigl\}\cdot\mathbf{E}\nonumber \\
&& =\mu_0\omega^2(\mathbf{P}_P^N+\mathbf{P}_B^N),
\end{eqnarray}
\begin{figure}
  \includegraphics[width=0.8 \columnwidth ]{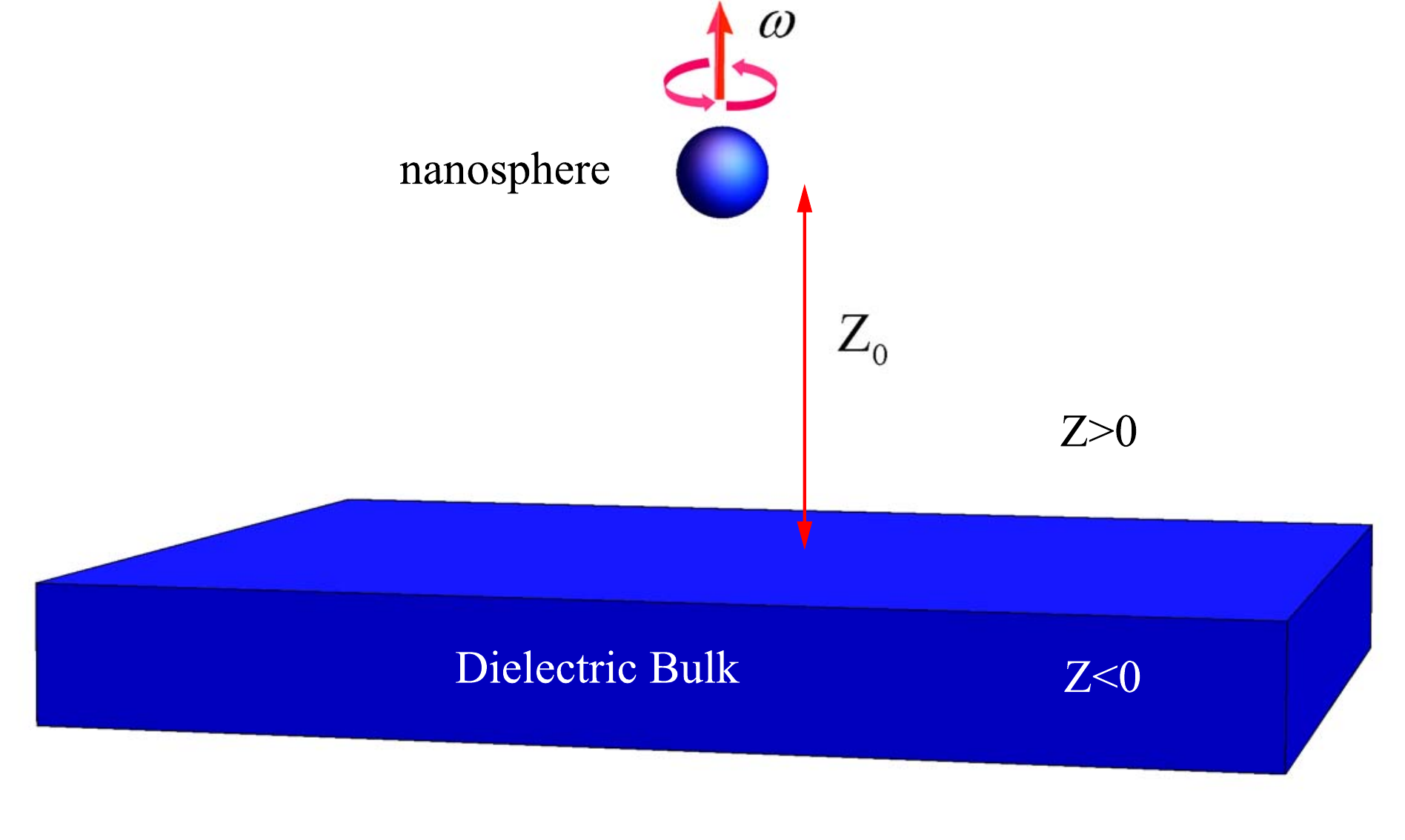}\\
  \caption{A rotating nanoparticle (with the radius $a=5nm$) above a semi-infinite dielectric (Bulk dielectric)}\label{setup}
\end{figure}
where $\mathbf{P}_P^N$ and $\mathbf{P}_B^N$ are the fluctuating or noise electric polarization components defined by
\begin{eqnarray}\label{noise}
P^N_{P,k} (\mathbf{r},t)=\epsilon_0\intf d\nu\,f_{ki}(\nu)\,X^N_i (\mathbf{r},\nu,t), \\ \nonumber
P^N_{B,k} (\mathbf{r},t)=\epsilon_0\intf d\nu\,g_{ki}(\nu)\,Y^N_i (\mathbf{r},\nu,t).
\end{eqnarray}
The fluctuating or noise matter fields $X^{N}$ and $Y^{N}$, are the homogeneous solutions of the equations of motion for the matter fields. One can expand these noise fields in terms of the ladder operators in the body frame as
\begin{eqnarray}\label{NField}
X^{N}_i (\rho,\varphi,z,\nu,t)&=&\sum_{m}\bigl[e^{i m\varphi} e^{i (\nu-m\omega_0)t}\,a^{\dag}_{i,m} (\rho,z,\nu)\\ \nonumber &+&e^{-i m\varphi}e^{-i (\nu-m\omega_0)t}\,a_{i,m} (\rho,z,\nu)\bigr], \\ \nonumber
Y^N_i(\mathbf{r},\nu,t)&=&e^{i\nu t}\,b_i^{\dag}(\mathbf{r},\nu)+e^{-i\nu t}\,b_i(\mathbf{r},\nu),
\end{eqnarray}
and from the canonical quantization rules Eq.(\ref{Q1})
\begin{eqnarray}\label{amamdag}
&&[a_{i,m} (\rho,z,\nu),a^{\dag}_{j,m'} (\rho',z',\nu')]=\frac{\hbar}{4\pi\nu}\,\delta_{ij}\,\delta_{mm'}\delta(\nu-\nu') \nonumber \\
&&\frac{\delta(\rho-\rho')\delta(z-z')}{\rho},  \nonumber \\
&&[b_{i} (\mathbf{r},\nu),b^{\dag}_{j} (\mathbf{r'},\nu')]=\frac{\hbar}{2\nu}\,\delta_{ij}\,\delta(\nu-\nu')\delta(\mathbf{r}-\mathbf{r'}).
\end{eqnarray}
If the dielectric bodies are held in temperature $T$, then
\begin{eqnarray}\label{FLUC-a}
&&\la a^{\dag}_{i,m} (\rho,z,\nu)\,a_{j,m'} (\rho',z',\nu')\ra_T=\frac{\hbar}{4\pi\nu}\,n_{T} (\nu)\,\delta_{mm'}\delta_{ij}\,\nonumber \\
&&\delta(\nu-\nu')\frac{\delta(\rho-\rho')\delta(z-z')}{\rho}, \\ \nonumber
&&\la b^{\dag}_{i} (\mathbf{r},\nu),b_{j} (\mathbf{r'},\nu')\ra_T=\frac{\hbar}{2\nu}\,n_{T}(\nu)\, \delta_{ij}\,\delta(\nu-\nu')\delta(\mathbf{r}-\mathbf{r'}),
\end{eqnarray}
where $n_{T}(\omega)=[\exp(\hbar\omega/k T)-1]^{-1}$.
\par In the body frame of the nanoparticle, the coupling tensor $f_{ij}$ is time independent and diagonal, corresponding to setting $\omega_0=0$. In this frame, the response functions denoted by $\chi^{0}_{kj}(\omega)$ can be obtained in terms of the diagonal components of the coupling tensor as \cite{kheirandish2014electromagnetic}
\begin{equation}\label{kapa-0}
\chi^{0}_{kk}(\omega) =\epsilon_0\intf d\nu\, \frac{f^{2}_{kk} (\nu)}{\nu^2-\omega^2},
\end{equation}
leading to
\begin{equation}\label{kapa-0}
\frac{f^2_{kk} (\nu)}{\nu}=\frac{2}{\pi\epsilon_0}\,\mbox{Im}[\chi^{0}_{kk} (\nu)].
\end{equation}
It can be easily shown that the response functions of the rotating nanoparticle in the laboratory frame and the body frame are related trough the following equations \cite{kheirandish2014electromagnetic}
\begin{eqnarray}\label{connection}
 \chi^{P}_{zz}(\omega,m)&=&\chi^{0,P}_{zz}(\omega-m\omega_0),\\ \nonumber
 \chi^{P}_{xx}(\omega,m)&=&\chi^{P}_{yy}(\omega,m) \\ \nonumber
 &=&\frac{1}{2}[\chi^{0,P}_{xx}(\omega_{+}-m\omega_0)+\chi^{0,P}_{xx}(\omega_{-}-m\omega_0)],\\ \nonumber
 \chi^{P}_{xy}(\omega,m)&=&-\chi^{P}_{yx}(\omega,m)
\\ \nonumber&=&\frac{1}{2i}[\chi^{0,P}_{xx}(\omega_{+}-m\omega_0)-\chi^{0,P}_{xx}(\omega_{-}-m\omega_0)].
\end{eqnarray}
For the non moving bulk dielectric the response functions in the body frame and the laboratory frame are the same.
\begin{figure}
    \includegraphics[width=1.0 \columnwidth]{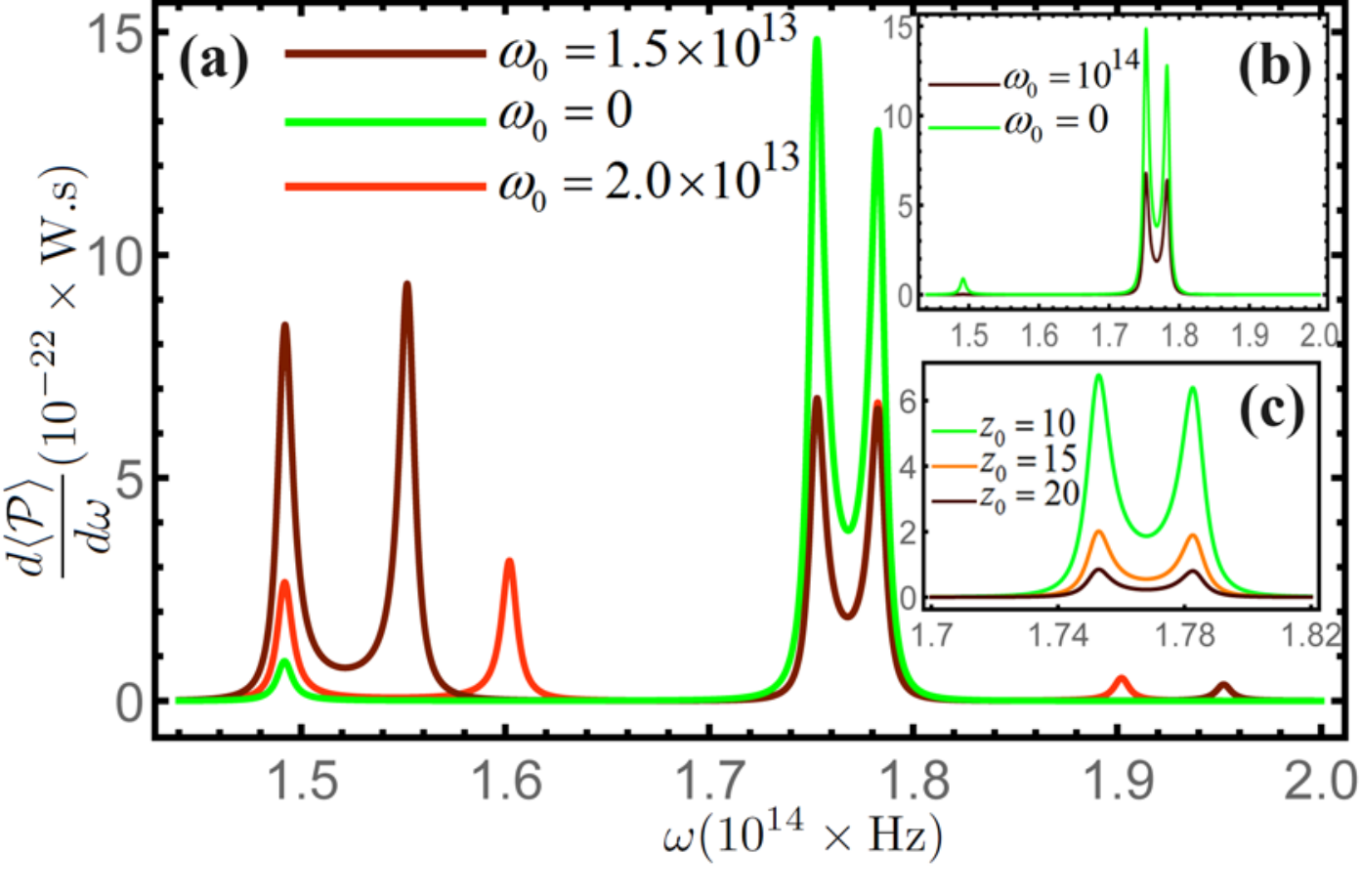}\\
  \caption{The absorbed radiation heat power spectrum as a function of frequency for different angular velocities (a,b) and for different distances from the surface $z_0(nm)$ (c). The bulk is hold at $T=300 k$ and for (a,b) the rotating nanoparticle is assumed to be at the distance $z_0=10 nm$ and is rotating with a constant angular velocity $\omega_0=10^{14}Hz$ in (c).}\label{det}
\end{figure}
\begin{figure}
    \includegraphics[width=1.0 \columnwidth]{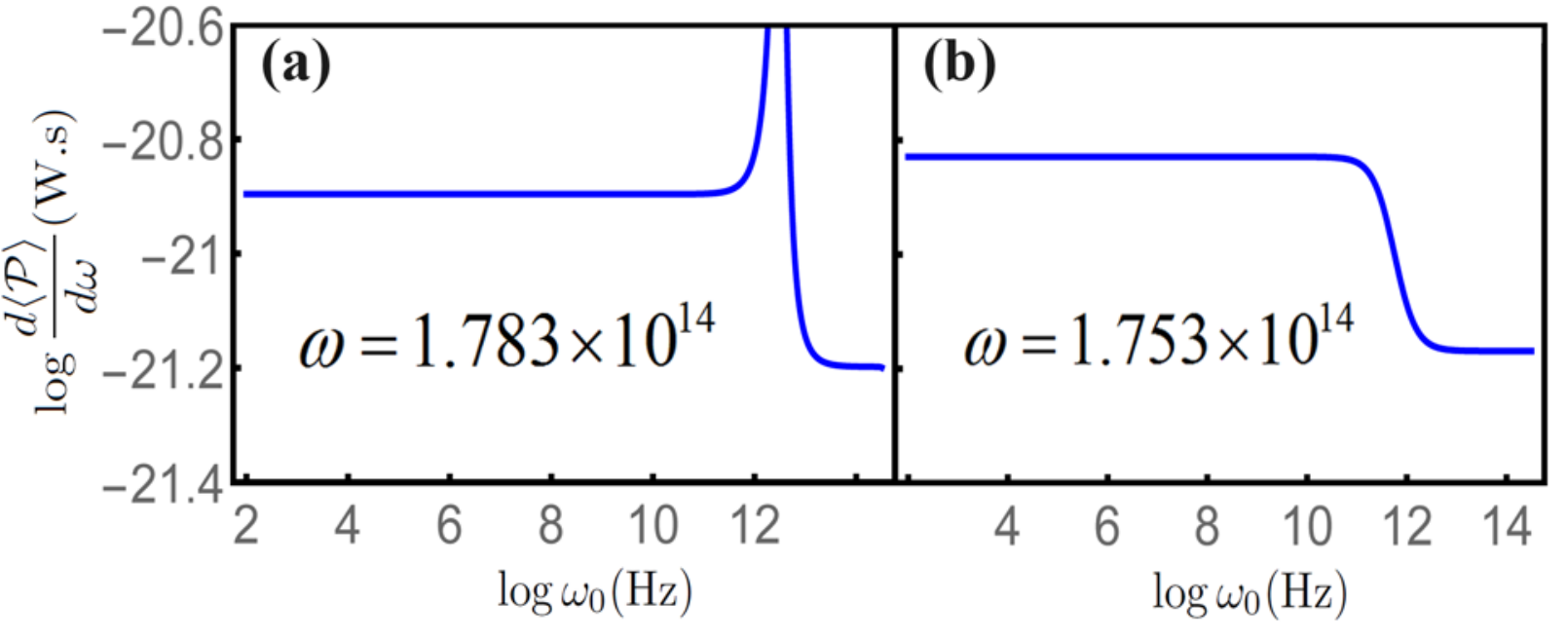}\\
  \caption{The logarithmic plot of the absorbed radiation heat power spectrum as a function of angular velocity of the nanoparticle for the peak frequencies a) $\omega=1.783\times 10^{14}$ b) $\omega=1.753\times 10^{14}$.}\label{dis}
\end{figure}
\begin{figure}
   \includegraphics[width=1.0 \columnwidth]{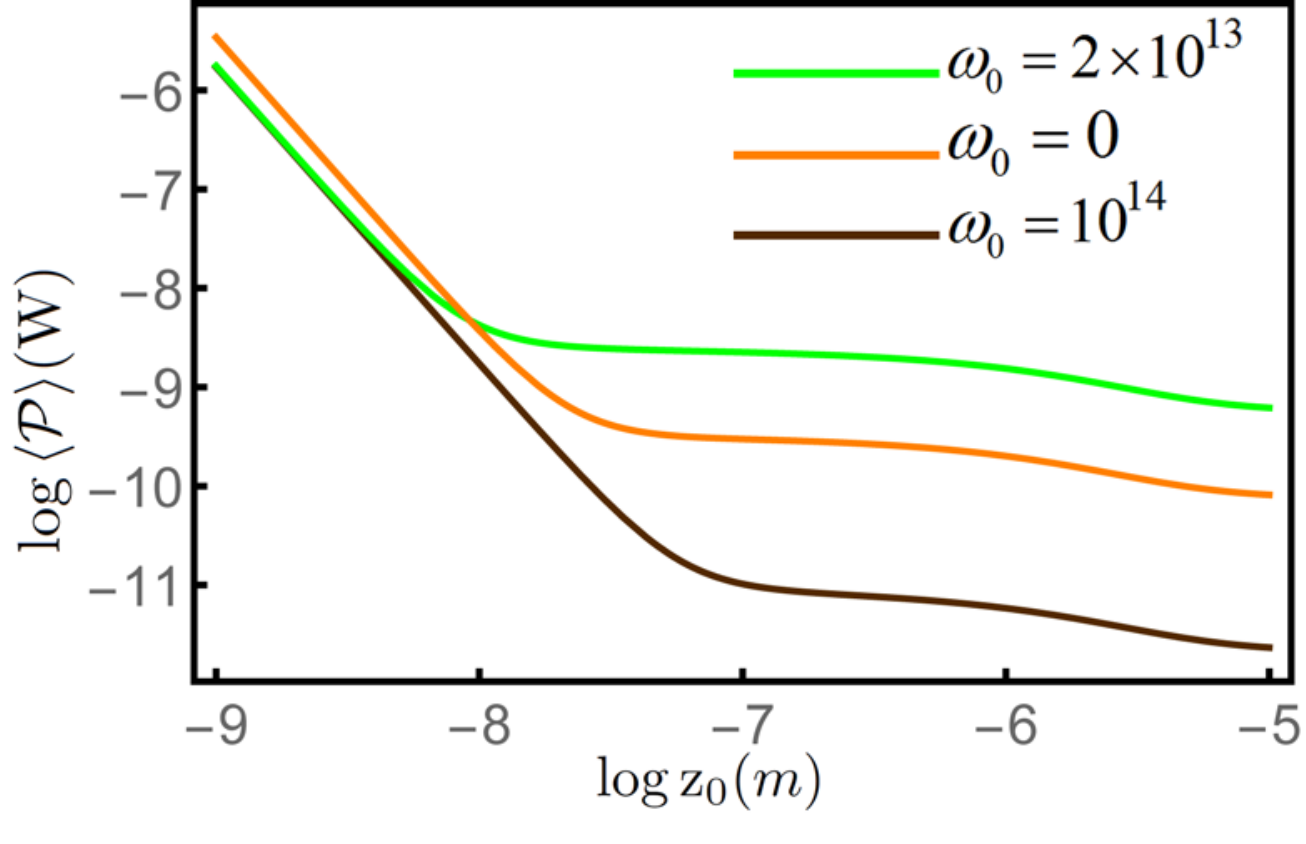}\\
  \caption{The logarithmic plot of the absorbed radiation heat power versus distance to the surface for different angular velocities.}\label{power}
\end{figure}
\par The rate of work done by the electromagnetic field on a differential volume $d\mathbf{r}$ of a dielectric is given by $\mathbf{j}\cdot (\mathbf{E}+\mathbf{v}\times\mathbf{B})\, d\mathbf{r}$, where $\mathbf{j}=\pt\mathbf{P}-\curl (\mathbf{v}\times\mathbf{P})$ is the current density in matter. In the non-relativistic regime, we can ignore from the terms containing the velocity $\mathbf{v}$, therefore, the radiated power of the rotating nanoparticle can be written as
\begin{equation}\label{Power}
 \la\mathcal{P}\ra=\int_{V} d\mathbf{r}\int\int_{-\infty}^{\infty}\frac{d\omega}{2\pi}\frac{d\omega'}{2\pi}e^{-i(\omega+\omega')t}
  (i\omega)\la \mathbf{P}(\mathbf{r},\omega)\cdot\mathbf{E}(\mathbf{r},\omega')\ra.
\end{equation}
Using the dyadic Green's tensor $G_{ij}$ and Eq.(\ref{FE}), we find
\begin{eqnarray}\label{e39}
E_i (\mathbf{r},\omega) &=& E_{0,i} (\mathbf{r},\omega)+\mu_0\omega^2\int d\mathbf{r}'\,G_{ij} (\mathbf{r},\mathbf{r}',\omega)\,\nonumber \\ && \times (P^{N}_{P,j} (\mathbf{r}',\omega)+P^{N}_{B,j} (\mathbf{r}',\omega)).
\end{eqnarray}
The scalar product $\la \mathbf{P}(\mathbf{r},\omega)\cdot\mathbf{E}(\mathbf{r},\omega')\ra$ in Eq.(\ref{Power}) leads to three non-zero expectation values $\la E_{0,i}(\mathbf{r},\omega)\cdot E_{0,j}(\mathbf{r'},\omega')\ra$, $\la P^N_{P,i}(\mathbf{r},\omega)\cdot P^N_{P,j}(\mathbf{r'},\omega')\ra$ and $\la P^N_{B,i}(\mathbf{r},\omega)\cdot P^N_{B,j}(\mathbf{r'},\omega')\ra$, where the first two terms have been studied as thermal and frictional radiation in \cite{kheirandish2014electromagnetic}, while the third term, which we will focus on in the following, is responsible for the heat transferred from the semi-infinite dielectric to the rotating nanoparticle.
\par Using Eqs.(\ref{FE},\ref{Power},\ref{FLUC-a}) and introducing $\mbox{Im}\alpha_{ij}(\omega)=V\mbox{Im}\chi^P_{ij}(\omega, m=0)$, we find the following general formula for the absorbed Radiation Heat power ($\la\mathcal{P}\ra_{RHT}$) for a rotating nanoparticle above a planar surface
\begin{eqnarray}\label{P1}
\la\mathcal{P}\ra_{RHT}&=& \frac{2\hbar}{\pi} \intf d\omega \frac{\omega^5}{c^4} \mbox{Im}\alpha_{ij} (\omega) \mbox{Im}\chi ^B(\omega) n_T(\omega)\nonumber \\ && \times\int d\mathbf{r'} G_{ik}(\mathbf{r},\mathbf{r'},\omega) G_{kj}^*(\mathbf{r},\mathbf{r'},\omega) .
\end{eqnarray}
Eq.(\ref{P1}) in the non rotating case $\omega_0=0$, tends to the result obtained in \cite{mulet2001nanoscale, shchegrov2000near} for a static nanoparticle above a surface hold at temperature $T$. The dyadic Green's functions $G_{ij}$ for this geometry have been calculated in details in \cite{maradudin1975scattering}.
\par To find some numerical results, let us consider a nanoparticle and a bulk dielectric made of silicon carbide (SiC) where the dielectric function is given by the oscillator model \cite{palik1998handbook},
\begin{equation}
\varepsilon(\omega)=\varepsilon_\infty(1+\frac{\omega_L^2-\omega_T^2 }{\omega_T^2-\omega^2-i\Gamma \omega }),
\end{equation}
with $\varepsilon_\infty=6.7$, $\omega_L=1.823\times 10^{14}$, $\omega_T=1.492\times 10^{14}$, and $\Gamma = 8.954\times 10^{11}$. Inserting the dyadic Green's functions given in \cite{maradudin1975scattering} into Eq.(\ref{P1}), the spectrum of the absorbed radiation heat power by the nanoparticle is plotted in different cases.
\par Fig.(\ref{det}) shows the spectrum of the absorbed radiation heat power by the rotating nanoparticle for some interesting angular velocities (a,b) and also for different distances from the surface (c). We realize that for angular velocities much smaller than the relaxation frequency ($\Gamma$) of the dielectrics ($\omega_0\ll\Gamma$), rotation can not affect the spectrum. In (a) and (b) the spectrum for $\omega_0>\Gamma$ and $\omega_0\gg\Gamma$ are depicted, respectively. As a result of (a) there are two remarkable peaks in agreement with the reported results for a static nanoparticle \cite{mulet2001nanoscale} and there are two more peaks which depend on the angular velocity of the nanoparticle, while in (b) we have peaks in the spectrum with smaller absorbed radiation heat power spectrum by the rotating nanoparticle. Increasing in absorbed radiation heat power as $1/z_0^3$, where $z_0$ is the distance from the surface and $\omega_0=10^{14}Hz$, is shown in (c). For the radiation heat transfer for a static nanoparticle see \cite{mulet2001nanoscale,dedkov2008vacuum}.
\par As mentioned before, the rotations with angular velocities much smaller than the relaxation frequency of the nanoparticle can not affect the spectrum. To illustrate this, let us consider the effect of rotation on the peaks of the spectrum ($\omega=1.753\times10^{14}$ and $\omega=1.783\times10^{14}$). In Fig.(\ref{dis}) the absorbed heat power $\frac{d\la\mathcal{P}\ra}{d\omega}$ is depicted for for peak frequencies ($\omega=1.753\times10^{14}$ and $\omega=1.783\times10^{14}$) in terms of the angular velocity $\omega_0$. The effect of rotation appears around $\Gamma(8.954\times 10^{11})$ and these plots resembling a phase transition, where after a sharp change around $\Gamma$, both plots get constant values. One may say, the rotation of the nanoparticle can cause a phase transition on absorbed radiation heat power and some interesting phenomena may appear near the critical frequency shown in Fig.(\ref{det})(a).
\par Fig.(\ref{power}) shows the total absorbed radiation heat power by the rotating nanoparticle as a function of the distance $z_0$ for some angular velocities. The absorbed power increases as $1/z_0^3$ at small distances from the surface known as the near field effect. However, the absorbed radiation heat  increases as $1/z_0^3$ just in distances less than $10nm$ for $\omega_0=2\times10^{13}$ while for $\omega_0=0$ and $\omega_0=10^{14}$, this behaviour remains valid up to distances $20nm$ and $50nm$, respectively. The results for $\omega_0=0$ are in agreement with those reported in \cite{mulet2001nanoscale}.
\begin{acknowledgments}
The authors wish to thank the graduate office of University of Isfahan for their support.
\end{acknowledgments}
\bibliography{sci}
\bibliographystyle{apsrev4-1}

\end{document}